# WHAT CAN PREDICTIVE SPEECH CODERS LEARN FROM SPEAKER RECOGNIZERS?[1]

Marcos Faúndez-Zanuy
Escola Universitaria Politècnica de Mataró (UPC)
Avda. Puig i Cadafalch 101-111
08303 MATARO (BARCELONA) SPAIN
tel: (34) 93 757 44 04 fax: (34) 93 757 05 24 e-mail: faundez@eupmt.es

## ABSTRACT

This paper compares the speech coder and speaker recognizer applications, showing some parallelism between them. In this paper, some approaches used for speaker recognition are applied to speech coding in order to improve the prediction accuracy. Experimental results show an improvement in Segmental SNR (SEGSNR).

## INTRODUCTION

This section describes the speech coding and speaker recognition application trying to find common facts on their formulations, and new potential possibilities with the cross-fertilization between both applications, and a joint formulation suitable for both of them.

### Similarities between speaker recognition and predictive speech coding

Several similarities exist between these applications. Among them, the most important are:

*Problem statement*

Without loss of generality we will assume that the speech signal $x(t)$ is normalized in order to achieve maximum absolute value equal to 1 ( $\max\left(\left|x(t)\right|\right)=1$). Otherwise, the following normalization can be done:

$$x'(t)=\frac{x(t)}{\max\left(\left|x(t)\right|\right)}$$

For speaker recognition applications, with a closed set of $N$ users inside the database, the problem is: given a vector of samples $\left[x(1),x(2),\cdots,x(L)\right]$, try to guess to whom speaker $speaker_i$ it belongs, with $speaker_i\in\left[1,2,\cdots,N\right]$. In order to achieve statistical consistency, this task is performed using several hundreds of vectors, and some kind of parameterization is performed over the signal samples, like a bank of filters, cepstral analysis, etc.

For speech coding applications, the problem is: given a vector of previous samples $\left[x(1),x(2),\cdots,x(L)\right]$, try to guess which is the next sample: $x(L+1),\quad x(L+1)\in\left[-1,1\right]$.

Thus, the problem statement is the same with the exception that the former corresponds to a discrete set of output values, and the latter corresponds to a continuum set of output values.

Taking into account this fact, the “speech predictor” can be seen as a “classifier”.

*Signal parameterization*

Although strong efforts have been done in speaker recognition for obtaining a good parameterization of the speech samples in order to improve the results and to reduce the computational burden, this step is ignored in speech coding, where the parameterized signal is the own signal without any processing.

*Model Computation*

For speaker recognition applications, some kind of model must be computed. Usually it is computed using some training material different from the test signals to be classified. This model [1] can be as simple as the whole set of input vectors (Nearest Neighbor model [2]) or the result of some reduction applied on them (Codebook obtained with Vector Quantization [3], Gaussian Mixture Model [4], etc.), being the most popular the GMM.

[1] This work has been supported by the Spanish grant CICYT TIC2000-1669-C04-02 and COST-277

For predictive speech coding applications, the model is usually a LPC (Linear predictive Coding) model, but it can also be a codebook [5], Volterra series [6], neural networks [7], etc. These last kind of predictors belong to the nonlinear prediction approaches, that can outperform the classical linear ones [8]. If the neural network is a Radial Basis Function [9], the similarity is considerable with the GMM model of the speaker recognition applications.

*Decision*

For speaker recognition applications, the decision (classification) is done taking into account the fitness of the test observation to the previous computed models, and some decision rule (for instance, maximum likelihood). On the other hand, in predictive speech coders, the predicted sample is the output of the predictor given one input vector. While the speech prediction is the result of one single predictor, in speaker recognition it is often used a combination of several classifiers performing their task over the same observations or sometimes even different (multimodal biometrics). This strategy is known as committee machines [10], and it can be considered that it relies on the principle of divide and conquer, where a complex computational task is split into several simpler tasks plus the combination of them. A committee machine is a combination of experts by means of the fusion of knowledge acquired by experts in order to arrive to an overall decision that it is supposedly superior to that attainable by any one of them acting alone.

## PROPOSALS

We propose two main improvements in predictive speech coding that. Both of them are well known in the context of speaker recognition.

*Contribution on "signal parameterization"*

One drawback (or perhaps advantage) of the classical LPC predictors is that there is just one parameter that can be set up: the prediction order (or vector dimension of the input vectors). Theoretically the higher the prediction order, the higher the accuracy of the prediction, but there is a saturation, even for high prediction orders. This fact can be seen looking at figure 1, that shows the SEGSNR of a predictive speech coder for the frame on the top of figure 1. The improvements at 40 and 80 prediction orders are due to the pitch periodicity of this frame, so a combined predictor with short term and pitch prediction schemes in order to achieve similar results with smaller prediction coefficients. This can be seen as a combination of predictors. Using a nonlinear predictor there is more feasibility and better results, because linear models are optimal just for gaussian signals, that it is not the case of speech signals. Another advantage of nonlinear models for (instance neural networks) is that they can integrate different kinds of information, they can use "hints" in order to improve the accuracy, etc. For this reason we propose, in addition to the speech samples, the use of delta parameters. This kind of information has been certainly useful for speaker recognition [11], where it is found that instantaneous and transitional representations are relatively uncorrelated, thus providing complementary information for speaker recognition. The computation of the transitional information is as simple as the first order finite difference. This transitional information is also known as delta parameters.

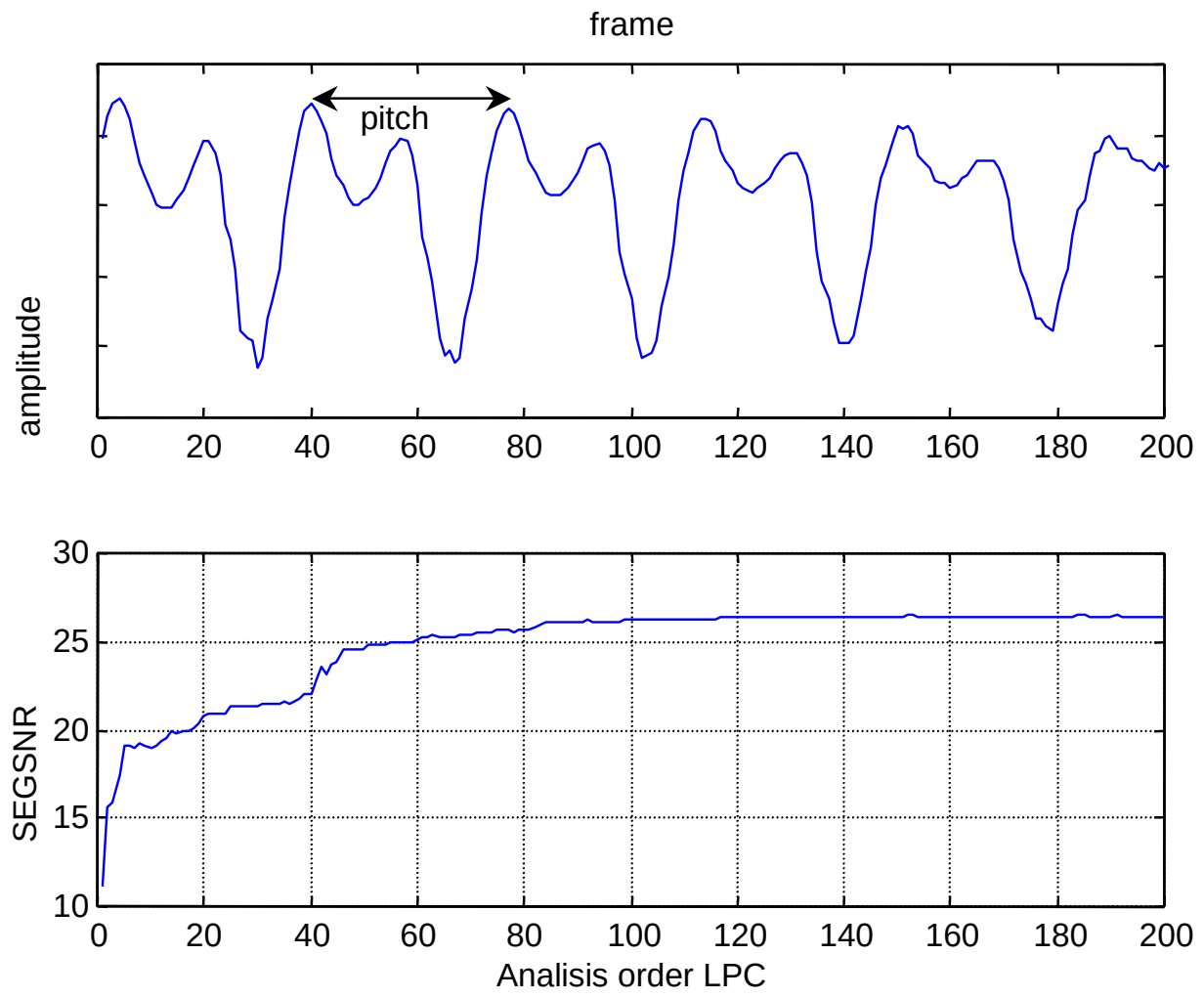

*Figure 1. SEGSNR vs prediction order.*

*Contributions on "decision"*

The combination of several predictors is similar to the Committee machines strategy [10]. If the combination of experts were replaced by a single neural network with a large number of adjustable parameters, the training time for such a large network is likely to be longer than for the case of a set of experts trained in parallel. The expectation is that the differently trained experts converge to different local minima on the error surface, and overall performance is improved by combining the outputs of each predictor. This approach was proposed and studied in [12], where a multistart algorithm was used in order to train the predictor. Thus, different predictors are combined, being the result of the same architecture, with the same training vectors and algorithm, but different initialization weights and biases. Another possibility, proposed in [13], consists on the combination of three different nonlinear predictors: Radial Basis Functions, Elman and Multi Layer Perceptron (MLP) neural networks.

## EXPERIMENTAL RESULTS

This section presents the conditions of the experiments and the experimental results.

*Database*

The experiments have been done using the same database of our previous work on nonlinear speech coding that can be found in [14]. We have encoded eight sentences uttered by eight different speakers (4 males and 4 females).

*Predictor based on Radial Basis Functions*

Although in our previous work the best result was achieved with MLP, and the RBF networks may require more neurons than MLP or Elman networks, they can be fitted to the training data with less time. For this reason, we will use a RBF as nonlinear predictor.

The RBF network consists on a Radial Basis layer of *S* neurons and an output linear layer. The output of $i^{th}$ Radial Basis neuron is $R_i = radbas\left(\|\vec{w}_i - \vec{x}\| \times b_i\right)$ , where:

- $\vec{x}$ is the *L* dimensional input vector
- $b_i$ is the scalar bias or spread (σ) of the gaussian
- $\vec{w}_i$ is the *L* dimensional weight vector of the Radial Basis neuron *i*, also known as center.
- The transfer function is $radbas[n] = e^{-n^2}$

In our case, the output is just one neuron. Figure 2 shows the scheme of a RBF network.

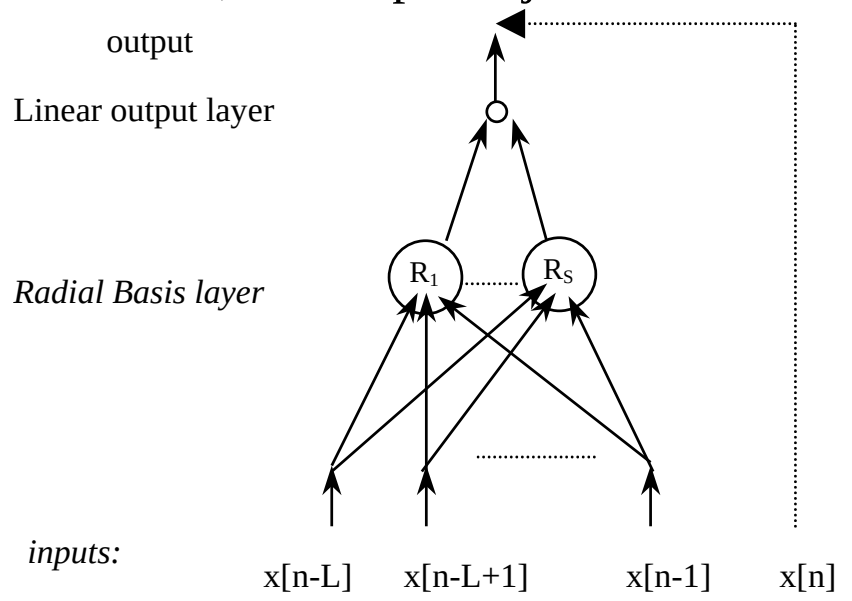

*Figure 2. RBF network architecture.*

The radial basis function has a maximum of 1 when its input is 0. As the distance between $\vec{w}$ and $\vec{x}$ decreases, the output increases. Thus, a radial basis neuron acts as a detector that produces 1 whenever the input $\vec{x}$ is identical to its weight vector $\vec{w}_i$. The bias *b* allows the sensitivity of the *radbas* neuron to be adjusted. For example, if a neuron had a bias of 0.1 it would output 0.5 for any input vector $\vec{x}$ at vector distance of 8.326 (0.8326/b) from its weight vector $\vec{w}_i$, because $e^{-0.8326^2} = 0.5$.

Radial basis neurons with weight vectors quite different from the input vector $\vec{x}$ have outputs near zero. These small outputs have only a negligible effect on the linear output neurons. In contrast, a radial basis neuron with a weight vector close to the input vector $\vec{x}$ produces a value near 1. If a neuron has an output of 1 its output weights in the second layer pass their values to the linear neurons in the second layer. In fact, if only one radial basis neuron had an output of 1, and all others had outputs of 0's (or very close to 0), the output of the linear layer would be the active neuron's output weights. This would, however, be an extreme case. Each neuron's weighted input is the distance between the input vector and its weight vector. Each neuron's net input is the element-by-element product of its weighted input with its bias.

---

We have used two different training algorithms:

<u>RBF-1:</u>

- The variance is manually setup in advance. Thus, it is one parameter to fix.
- The algorithm iteratively creates a radial basis network one neuron at a time. Neurons are added to the network until the maximum number of neurons has been reached.
- At each iteration the input vector that results in lowering the network error the most, is used to create a radial basis neuron.

<u>RBF-2:</u>

- The weights are all initialized with a zero mean, unit variance normal distribution, with the exception of the variances, which are set to one.
- The centers are determined by fitting a Gaussian mixture model with circular covariances using the EM algorithm. (The mixture model is initialized using a small number of iterations of the K-means algorithm). The variances are set to the largest squared distance between centers
- The hidden to output weights that give rise to the least squares solution are determined using the pseudo-inverse.

<u>*Predictive speech coding*</u>

We have used an ADPCM scheme with an adaptive scalar quantizer based on multipliers [15]. The number of quantization bits is variable between Nq=2 and Nq=5, that correspond to 16 kbps and 4 0kbps (the sampling rate of the speech signal is 8 kHz). We have used a prediction order $L$=10

<u>Experiments with algorithm RBF-1</u>

In order to setup the RBF architecture, we have studied the relevance of two parameters: spread (variance) and number of neurons. First, we have evaluated the SEGSNR for an ADPCM speech coder with RBF prediction and adaptive quantizer of 4 bits, as function of the spread of the gaussian functions. Figure 3, on the left, shows the results using one sentence, for spread values ranging 0.011 to 0.5 with an step of 0.01 and $S$=50 neurons. It also shows a polynomial interpolation of third order, with the aim to smooth the results. Based on this plot, we have chosen a spread value of 0.22. Using this value, we have evaluated the relevance of the number of neurons. Figure 3, on the center, shows the results using the same sentence and a number of neurons ranging from 5 to 100 with an step of 5. This plot also shows an interpolation using a third order polynomial. Using this plot we have chosen an RBF architecture with $S$=20 neurons. If the number of neurons (and/ or the spread of the guassians) is increased, there is an overfit (over parameterization that implies a memorization of the data and a loose of the generalization capability). In order to check the consistency of this settings, we have evaluated again the results with 20 neurons and variable spread ranging between 0.011 and 1.2 with an step of 0.01. Figure 1 on the right shows the results and a fourth order polynomial interpolation. It is clear comparing the left and right plots of figure 3, that the spread should be increased to 0.37. This is a reasonable result, because if the number of neurons is reduced from 50 to 20, the spread should be increased in order to cover the “holes” left free by the removed centers.

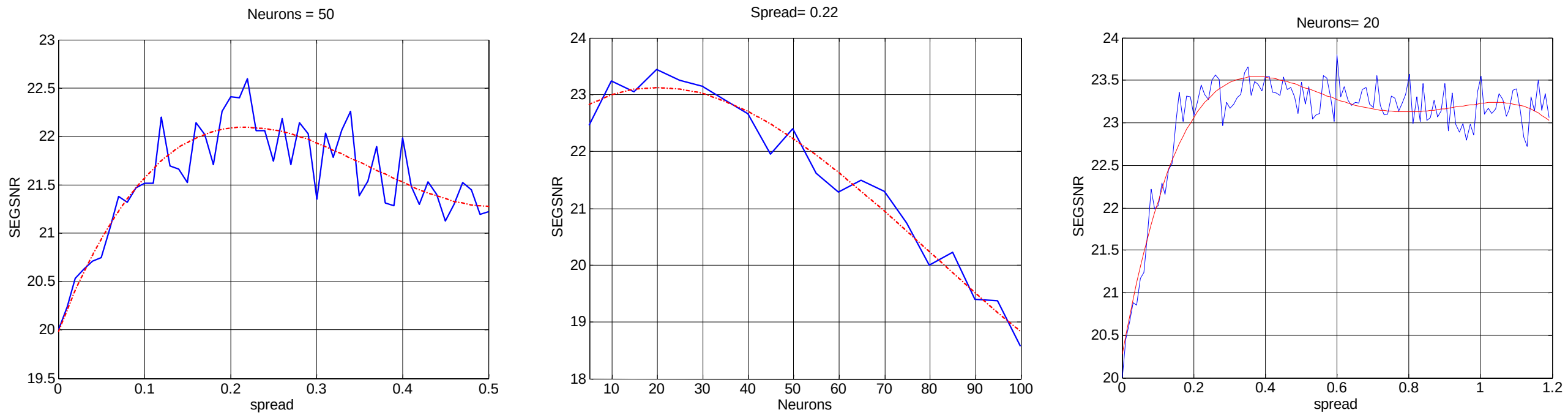

*Figure 3. RBF-1, left: SEGSNR vs spread for 50 neurons; center: SEGSNR vs number of neurons for spread=0.22: right: SEGSNR vs spread for 20 neurons.*

The use of delta parameters implies the replacement of the RBF architecture of figure 2 by the scheme shown in figure 4, were the delta parameters are computed with the following formulation:

Given the pair of (*inputs, output*) values used to train the RBF in the classical by the expression:

$$\left(\underbrace{\left[x(1), x(2), \cdots, x(L)\right]}_{inputs}, \underbrace{x(L+1)}_{output}\right) = \left(\vec{x}_1, x(L+1)\right)$$

$$\left(\left[x(2), x(3), \cdots, x(L+1)\right], x(L+2)\right) = \left(\vec{x}_2, x(L+2)\right)$$

$$\cdots$$

$$\left(\left[x(1+N), x(2+N), \cdots, x(L+N)\right], x(L+N+1)\right) = \left(\vec{x}_{1+N}, x(L+N+1)\right)$$

we propose the use of delta information computed in the following way:

$$\vec{\Delta}_i = \left[\Delta(i), \Delta(i+1), \cdots, \Delta(L+i-1)\right]$$

$$\vec{\Delta}_i = \vec{x}_i - \vec{x}_{i-1} = \left[x(i) - x(i-1), x(i+1) - x(i), \cdots, x(L+i-1) - x(L+i-2)\right]$$

In this case, the input-output relation is:

$$\left(\left[\underbrace{\Delta(i), \Delta(i+1), \cdots, \Delta(L+i-1)}_{\text{Transitional information}}, \underbrace{x(i), x(i+1), \cdots, x(L+i-1)}_{\text{Instantaneous information}}\right], x(L+i)\right) = \left(\underbrace{\left[\vec{\Delta}_i, \vec{x}_i\right]}_{inputs}, \underbrace{x(L+i)}_{output}\right)$$

It is interesting to observe that this scheme just implies the addition of one more input sample, that is $x(L+i-2)$, when compared against the original system without transitional information. Thus, the predictor order is L+1, when using transitional information. This is due to the transitional information is obtained using a linear combination of samples already used in the instantaneous portion of the input. Obviously this won't imply an improvement in a linear predictor, but differences can be expected when entering a nonlinear predictor, because the squared value of a linear combination generates cross-product terms. Although it is supposed that the neural net is able to produce these terms by his own non-linearity, if they are explicitly entered, there is a reinforcement.

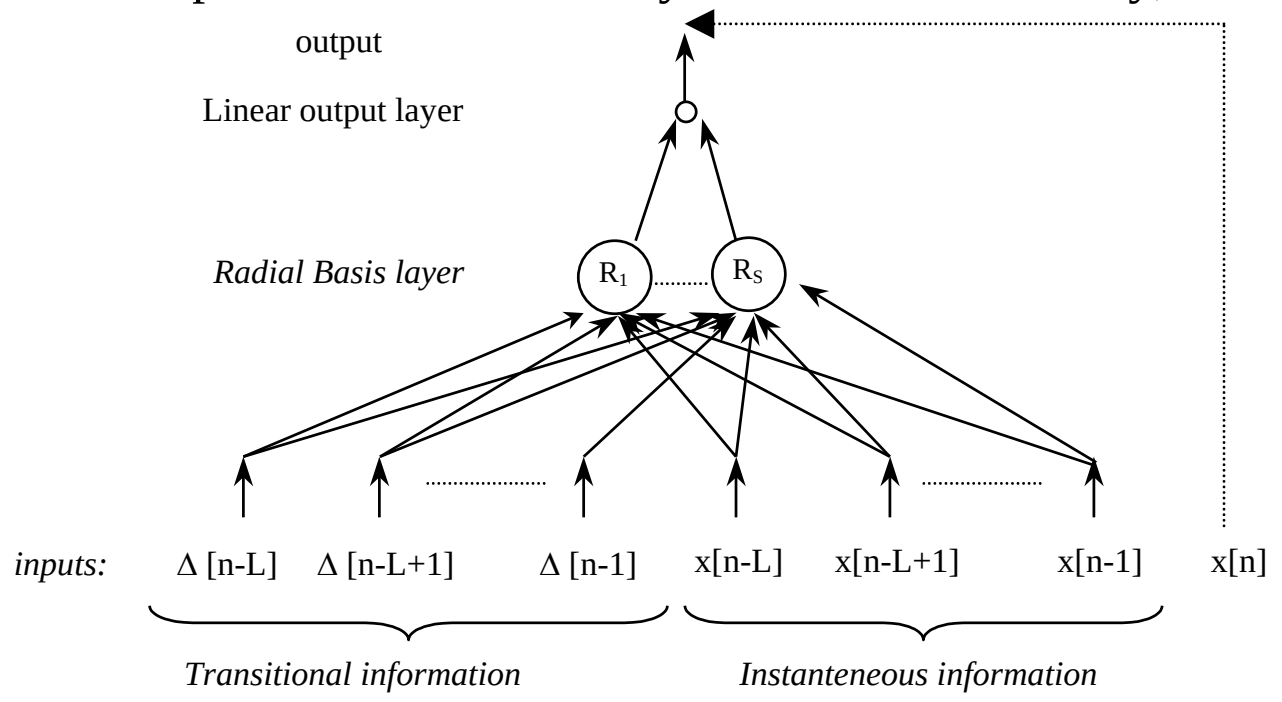

*Figure 4. RBF scheme using delta parameters*

Theoretically the scheme of figure 4 implies higher dimensional vectors (the number of inputs has been doubled!), so more centers and/ or spread should be necessary. However, figure 5 shows the results in the same conditions than figure 3 using the RBF scheme shown in figure 4. Comparing figures 3 and 5 it is clear that the use of delta information can improve the SEGSNR, even without an increase of the number of centers. However, the best results are achieved with an increase in spread from 0.22 to 0.4.

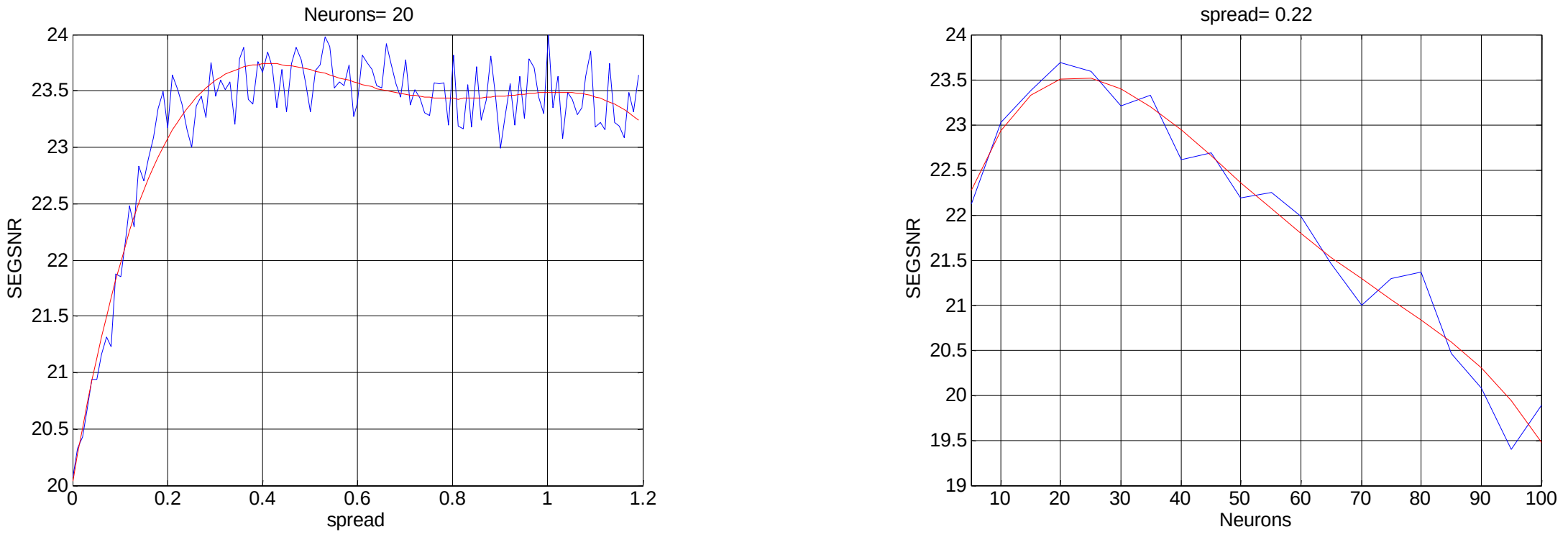

*Figure 5. RBF-1, delta parameters left: SEGSNR vs spread for 20 neurons; right: SEGSNR vs neurons for spread=0.22.*

<u>Experiments with algorithm RBF-2</u>

Comparing figures 3 and 5 it can be seen that the system with delta parameters achieves and improvement without an increase of the number of neurons (centers), for the whole range of spreads and number of neurons. Anyway, the algorithm needs the setup of the spread. In order to avoid this, we can use the algorithm RBF-2 described before, that overcomes this drawback.

Figure 6 shows the obtained results for RBF-2 and the architecture of figure 2.

Table 1 shows the results for RBF-1 (with spread=0.4), RBF-2 and a committee RBF-1+RBF-2 with and without delta parameters.

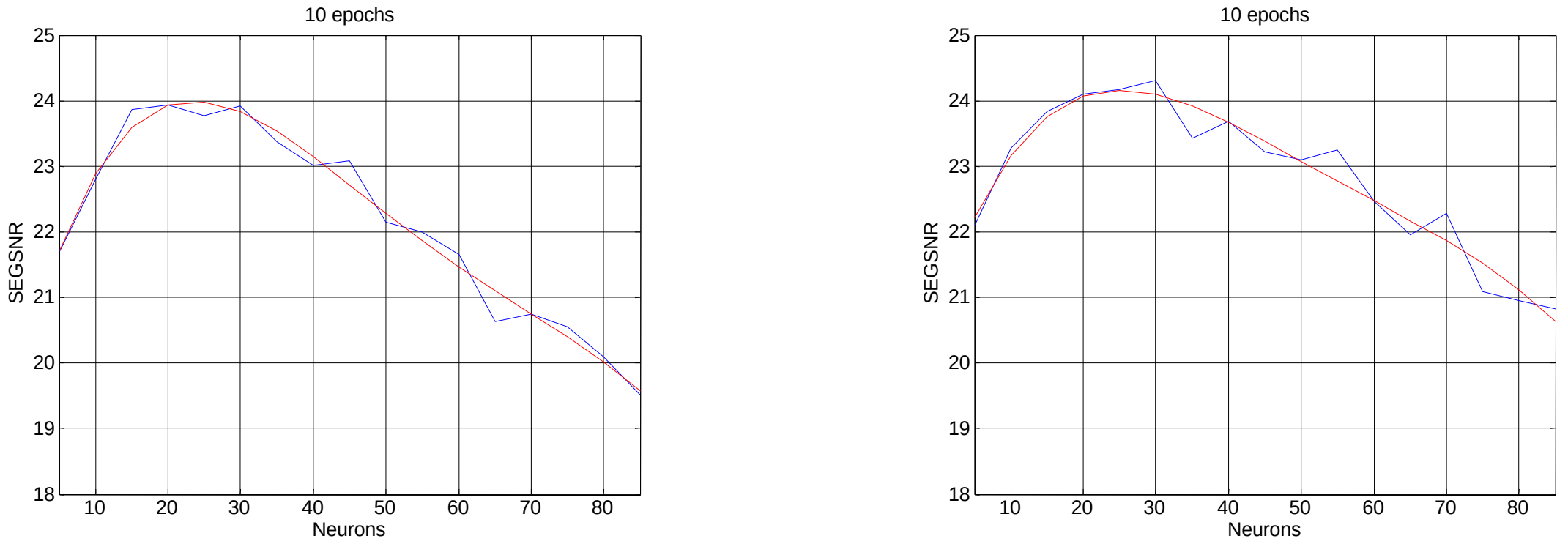

*Figure 6 SEGSNR vs number of neurons for RBF-2 trained with 10 epochs. Left: without delta parameters, right: with delta parameters.*

*Table 1. SEGSNR for ADPCM with several predictors.*

| | RBF-1 (spread=0.22) | | | | RBF-1 (spread=0.4) | | | | RBF-2 | | | |
|---|---|---|---|---|---|---|---|---|---|---|---|---|
| Parameterization → | $\vec{x}$ | | $\vec{x}+\vec{\Delta}$ | | $\vec{x}$ | | $\vec{x}+\vec{\Delta}$ | | $\vec{x}$ | | $\vec{x}+\vec{\Delta}$ | |
| Nq ↓ | m | σ | m | σ | m | σ | m | σ | m | σ | m | σ |
| 2 | 11.65 | 7.63 | 10.94 | 7.21 | 12.05 | 8.88 | 12.21 | 8.36 | 13.75 | 5.96 | 14.05 | 5.78 |
| 3 | 18.40 | 6.56 | 18.99 | 6.13 | 19.18 | 7.94 | 19.17 | 7.71 | 20.23 | 6.41 | 20.60 | 6.22 |
| 4 | 23.69 | 6.12 | 23.40 | 6.05 | 24.33 | 7.20 | 24.47 | 6.99 | 25.35 | 6.57 | 25.69 | 6.63 |
| 5 | 28.22 | 6.34 | 28.13 | 6.14 | 29.16 | 7.28 | 29.29 | 6.95 | 30.22 | 6.90 | 30.42 | 6.86 |

*Table 2. SEGSNR for ADPCM with several combined predictors.*

| | Combined RBF-1 spread=0.22+RBF2 | | | | Combined RBF-1 spread=0.4+RBF2 | | | |
|---|---|---|---|---|---|---|---|---|
| Parameterization → | $\vec{x}$ | | $\vec{x}+\vec{\Delta}$ | | $\vec{x}$ | | $\vec{x}+\vec{\Delta}$ | |
| Nq ↓ | m | σ | m | σ | m | σ | m | σ |
| 2 | 13.46 | 6.58 | 13.66 | 6.12 | 13.57 | 7.27 | 13.88 | 6.90 |
| 3 | 19.97 | 6.13 | 20.05 | 5.80 | 20.21 | 7.02 | 20.60 | 6.56 |
| 4 | 25.19 | 6.18 | 25.15 | 6.00 | 25.52 | 6.87 | 25.69 | 6.80 |
| 5 | 29.91 | 6.43 | 29.87 | 6.35 | 30.22 | 7.03 | 30.42 | 6.92 |

## CONCLUSIONS

A parallelism between speaker recognition and predictive speech coding has been established. This way of seeing has let us to propose two main improvements:

- The use of other information such as delta parameters.
- The combination between different predictors (or the same kind of predictor trained in a different way)

Both improvements can be applied to other speech coders.

This study reveals the potential of this strategies, and the possibility to establish another parallelisms and improvements.